\documentclass[pre,superscriptaddress,showkeys,showpacs,twocolumn]{revtex4-1}
\usepackage{graphicx}
\usepackage{latexsym}
\usepackage{amsmath}
\begin {document}
\title {Critical behaviour of a tumor growth model -- \\ Directed Percolation with a mean-field flavour}
\author{Adam Lipowski}
\affiliation{Faculty of Physics, Adam Mickiewicz University, Pozna\'{n}, Poland}
\author{Ant\'onio  Luis Ferreira}
\affiliation{Departamento de  Fisica and I3N, Universidade de Aveiro, 3810-193 Aveiro, Portugal}
\author{Jacek Wendykier}
\affiliation{Institute of Physics, Opole University, 45-052 Opole, Poland}
\begin {abstract}
We examine the critical behaviour of a lattice model of tumor growth where supplied nutrients are correlated with the distribution of tumor cells. Our results support the previous report (Ferreira et al., Phys. Rev. E {\bf 85}, 010901 (2012)), which suggested that the critical behaviour of the model differs from the expected Directed Percolation (DP) universality class. Surprisingly, only some of the critical exponents ($\beta$, $\alpha$, $\nu_{\perp}$, and $z$) take non-DP values while some others ($\beta'$, $\nu_{||}$, and spreading-dynamics exponents $\Theta$, $\delta$, $z'$) remain very close to their DP counterparts. The obtained exponents satisfy the scaling relations $\beta=\alpha\nu_{||}$, $\beta'=\delta\nu_{||}$, and the generalized hyperscaling relation $\Theta+\alpha+\delta=d/z$, where the dynamical exponent $z$ is, however, used instead of the spreading exponent~$z'$. Both in $d=1$ and $d=2$ versions of our model, the exponent $\beta$ most likely takes the mean-field value $\beta=1$, and we speculate that it might be due to the roulette-wheel selection, which is used  to choose the site to supply a nutrient.
\end{abstract}
\pacs{05.70.Fh, 87.18.Hf} \keywords{tumor growth, absorbing states, critical behaviour}

\maketitle
\section{Introduction}
Recently, critical behaviour of models with absorbing states is attracting a lot of interest \cite{hinrichsen,malte1}. A lot of such models appeared mainly due to a large number of phenomena where they are potentially applicable, such as, for example, in spreading of fires or epidemics, heterogeneous catalysis or granular flows.  Yet another motivation comes from the intriguing idea  that critical behaviour of these nonequilibirum systems can be, at least to some extent, categorized and understood in an analogous way  to equilibrium systems, where the notions of scale invariance, universality,  and renormalization group turned out to be so fruitful and productive~\cite{malte}. 

A prominent and strong indication that the universality is indeed common also among nonequilibrium systems comes with Directed Percolation (DP). Defined in a simple mathematical way, this spreading process turns out to exhibit the critical behaviour, which is very robust and appears in a number of other models as well as in some experimental realizations \cite{chate}.
A common feature of systems exibiting the DP criticality is that they have a single (or effectively single) absorbing state. The ubiquity of the DP critical behaviour prompted three decades ago Janssen \cite{janssen} and Grassberger \cite{grass} to formulate  a conjecture that all continuous transitions in systems with a single absorbing state and lacking extra symmetry or conservation law will belong to the DP universality class. Up to now, there are plenty of numerical examples supporting this conjecture \cite{hinrichsen,malte1}. Models with multiple absorbing states might also belong to the DP universality class \cite{jensen} but in some cases they exhibit a different behaviour \cite{lipdroz}. 

Absorbing states appear also in some ecological  or tumor growth models, where they correspond to the extinction of one or more species or types of cells. In some cases the critical points in these models were shown to belong to the Directed Percolation universality class \cite{rozenfeld,liplip,mobilia,wendykier}. However, an implementation of biologically relevant factors, such as, for example, inter- or intra-species competition, food supply, aging, birth or death, results in rather complex dynamics. Hence, a violation of the Grassberger and Janssen conjecture  cannot be excluded in such models.     

In the present paper, we examine the critical behaviour of a recently introduced tumor growth model \cite{ferliplip}. Our previous report \cite{ferliplip} suggested that despite a single absorbing state, the critical behaviour of this model differs from the expected DP universality class. Having performed much more extensive and detailed analysis, in the present paper we confirm the non-DP criticality of the model. However, the apparent violation of the Grassberger and Janssen conjecture occurs in a very intricate way: only some exponents take non-DP values while some others, including  all spreading-dynamics exponents, coincide with their DP counterparts. Nevertheless, some scaling and hyper-scaling relations seem to be satisfied in our model. 

\section{Model and numerical methods}

\subsection{Definition of the model}
In our model, each site of a $d$-dimensional lattice of linear size $N$  either is occupied by tumor, nutrient, both tumor and nutrient, or is empty. At a rate $p$, nutrients are supplied to a chosen site of the lattice, provided that the site is not already occupied by a nutrient. The roulette-wheel selection~\cite {liproulette} is used to choose the site for such a supply and the corresponding weight~$w$ depends on whether the site is occupied by tumor ($w=1+\Delta$) or not ($w=1$). The parameter $\Delta>0$ takes into account the angiogenic effects of increased nutrients supply due to formation of new blood vessels in the vicinity of tumor cells.
At a rate $1-p$, a tumor cell is updated on a randomly chosen site. The tumor cell survives provided that there is a nutrient on this site, otherwise it dies. The surviving tumor cell consumes the nutrient and attempts to breed provided that there is a site without a tumor cell among its nearest neighbours. Some preliminary results on this model were already reported  \cite{ferliplip}. Closely related models but without any preference in nutrient supply ($\Delta=0$) were also studied~\cite{wendykier}.

\subsection{Monte Carlo simulatons}
To examine our model, we used Monte Carlo simulations. 
To facilitate more extensive simulations, we implemented an algorithm where a site is selected only from among active sites, i.e., sites that are either without nutrients or with tumor cells. Maintaining the lists of such sites requires some bookkeeping but it speeds up the simulations substantially, especially in the vicinity of the critical point (where most sites are inactive). We checked that the simpler algoritm that does not keep the list of active sites and the present algorithm yield statistically the same results with respect both to static as well as to dynamic characteristics.

For various~$p$ and~$\Delta$, we measured the steady-state densities of tumor cells~$x_t$ and of nutrients~$x_n$. Simulations started from a random initial configuration and the model relaxed until a steady state was reached. To ensure that the obtained results are $N$-independent, we examined lattices of various sizes. Assuming that $x_t$ is an order parameter, we expected that upon approaching the critical point ($p=p_c$), it decays as $x_t\sim (p-p_c)^{\beta}$, where $\beta$ is a characteristic exponent. We also measured the time dependence of the tumor cell density~$x_t(t)$ (with the unit of time corresponding to~$N^d$ update attempts). The density $x_t(t)$ for each $t$ is an average over independent runs. At the critical point, the density~$x_t(t)$ is expected to have a power-law decay $x_t(t)\sim t^{-\alpha}$, where~$\alpha$ is another characteristic exponent~\cite{hinrichsen,comment1}. 

Some other exponents can be extracted assuming the following scaling form of the tumor cell density \cite{hinrichsen}
\begin{equation}
x_t(t)\sim t^{-\alpha}f((p-p_c)t^{1/\nu_{||}}, t^{d/z}/N)
\label{scaling-form}
\end{equation}
In particular, from the finite-size scaling analysis of the density $x_t(t)$ at the critical point ($p=p_c$), we can estimate the dynamical exponent $z$ from the collapse of the suitably rescaled $x_t(t)$ for various $N$. With a similar procedure, the exponent $\nu_{||}$ which describes the divergence of the temporal correlation length ($\xi_{||}\sim (p-p_c)^{-\nu_{||}}$),  can be  estimated from the off-critical steady-state data.

The divergence of the spatial correlation length $\xi_{\perp}$ is characterized with the exponent $\nu_{\perp}$ ($\xi_{\perp} \sim (p-p_c)^{-\nu_{\perp}}$). For finite-size systems at criticality this exponent governs also the vanishing of the steady-state order parameter \cite{jensen94}
\begin{equation}
x_t \sim N^{-\beta/\nu_{\perp}}
\label{nu-perp}
\end{equation}

Further information about the critical behaviour of our model can be obtained using the so-called spreading dynamics. In this technique one starts the simulation from a nearly-absorbing configuration and monitors the characteristics of the runs. 
In the case of our model, we started the simulations from the configuration where one site contains both a nutrient and a tumor cell and all other sites are filled only with nutrients.
We measured the average number (not density) of tumor cells ($X_t$), the survival probability $P(t)$ that a given run survives  until time $t$, and the mean square of spreading of tumor cells from the origin $R^2(t)$. 
At criticality one expects that
\begin{equation}
X_t(t)\sim t^{\Theta}, \ \ P(t)\sim t^{\delta}, \ \ R^2(t) \sim t^{2/z'}
\label{spreading}
\end{equation}
where $\Theta$, $\delta$ and $z'$ are characteristic exponents of the spreading process. Moreover, from the decay of the long-time limit of the survival probability 
\begin{equation}
P(t\rightarrow\infty)\sim (p-p_c)^{\beta'}
\label{betaprime}
\end{equation}
we can extract yet another characteristic exponent $\beta'$.

There are a number of examples showing that for models with a single absorbing state, the dynamical exponent $z$ and the spreading exponent $z'$ are equal. 
A violation of $z=z'$ equality was reported in some nonequilibrium Ising models but it was attributed to the missing absorbing phase \cite{nora2003}.
Moreover, for models with multiple absorbing states, the exponent~$z'$ might depend on the absorbing state that is chosen in spreading-dynamics simulations \cite{mendes94}.

The above critical exponents are not entirely independent. For models like Directed Percolation or Contact Process $\alpha=\delta$, $z'=z$, $\beta=\beta'$ and some plausible scaling arguments can be used to derive the following hyperscaling relation \cite{grassberger}
\begin{equation}
\Theta + 2\delta= \frac{d}{z}
\label{hyper}
\end{equation}
For models with multiple absorbing states the hyperscaling relation is expected to hold in a more general version \cite{mendes94}
\begin{equation}
\Theta + \alpha + \delta= \frac{d}{z'}
\label{ghyper}
\end{equation}
and in general $\beta\neq \beta'$.

Using scaling assumptions one can also show that \cite{hinrichsen}
\begin{equation}
\beta=\alpha\nu_{||},\ \ \beta'=\delta\nu_{||}.
\label{alphabeta}
\end{equation}
Moreover, the dynamical exponent $z$, which is defined from the relation $\xi_{||}\sim \xi_{\perp}^z$, must of course satisfy the equality
\begin{equation}
z=\nu_{||}/\nu_{\perp}.
\label{znunu}
\end{equation}

\subsection {Mean-field approximation}
To supplement Monte Carlo simulations, we investigated the behaviour of our model using the Mean-Field Approximation.
The general strategy to develop such an approach is to write a master equation and then solve it using some
decoupling procedure \cite{tome94,alf93,wendykier}. In the first step one sums the states of all but one site. As a result, one
obtains equations describing the time evolution of probabilities
for single-site configurations. Such equations will contain
probabilities of more complex configurations (i.e., two-site
configurations), which need to be  factorized subsequently
(i.e., approximated with some products of probabilities of
single-site configurations). Eventually, each term contributing
to the evolution of single-site probabilities will contain the probability
that a chosen site is in an appropriate state (for a given
process) and the rate of that process. In some
processes, neighboring sites also give some contributions. 
Using the following notation for a probability of a site being in particular state
\vspace{-1mm}
\begin{displaymath}
\begin{array}{ll}
x_1 - & {\rm empty}\\
x_2 - & {\rm nutrient\ only}\\
x_3 - & {\rm tumor\ only}\\
x_4 - & {\rm both\ nutrient\ and\ tumor}\\
\end{array}
\label{notation}
\end{displaymath}
 one arrives at the following equations describing
the time evolution of the probabilites (densities)~$x_i$, $i=1,2,3,4$:
\begin{eqnarray}
\frac{dx_1}{dt} & = &  (1-p)(x_3-fx_1x_4)-pgx_1 \nonumber \\
\frac{dx_2}{dt} & = &  pgx_1-(1-p)fx_2x_4 \nonumber  \\
\frac{dx_3}{dt} & = &  (1-p)(x_4-x_3+fx_1x_4)-pg(1+\Delta)x_3 \nonumber  \\
\frac{dx_4}{dt} & = &  (1-p)(fx_2x_4-x_4)+pg(1+\Delta)x_3, 
\label{mfa}
\end{eqnarray}
where $f=\frac{1-(x_3+x_4)^{2d}}{x_1+x_2}$ and $g=\frac{1}{1+\Delta (x_3+x_4)}$. 

Solving numerically the above set of equations, we determined the single-site probabilites and thus the tumor ($x_t=x_3+x_4$) and the nutrient ($x_n=x_2+x_4$) densities. To control the accuracy of the numerical procedure, one can check whether the obtained probabilites satisfy the obvious normalization condition $x_1+x_2+x_3+x_4=1$.

\section{Numerical results}

\subsection{d=1}
First, we describe the results for the $d=1$ version of the model. The already reported results of simulations show \cite{ferliplip} that, for a sufficiently large~$p$, the model remains in an active phase with $x_t>0$, which terminates at a critical point~$p_c$ (which depends on~$\Delta$). For $p<p_c$, the steady state of the model  is an absorbing state $x_t=0$ and $x_n=1$ (tumor cells die out and the system gets filled with nutrients).  At $p=p_c$, the model undergoes a phase transition from an active into an absorbing phase and we expect that $x_t$ is the corresponding order parameter. As it is already known, models with a single absorbing state are expected to belong to the so-called directed-percolation~(DP) universality class~\cite{hinrichsen}.  In this universality class and for $d=1$, the critical exponent $\alpha_{DP}=0.159(1)$ and the decay of the order parameter upon approaching the critical point is described by the exponent $\beta_{DP}=0.276(1)$. Numerical simulations show \cite{ferliplip} that for $\Delta=0$ (i.e., with no preference for placing the nutrients) the model  belongs indeed to the DP universality class. Much different behaviour, however, was reported for $\Delta>0$, where Monte Carlo simulations give $\beta=1.0(1)$ and $\alpha=0.60(5)$.

\begin{figure}
\includegraphics[width=9cm]{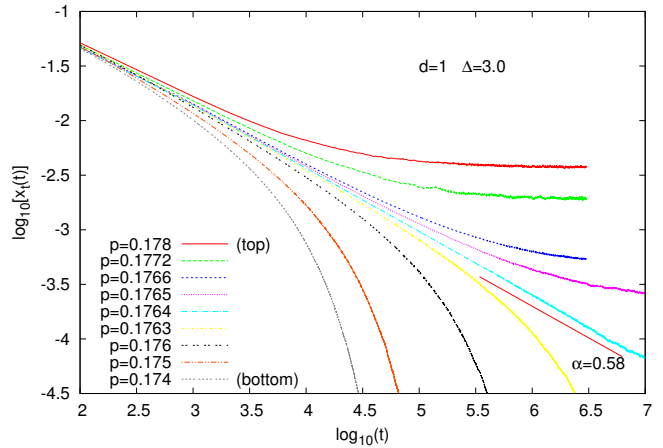}  \vspace{-0.7cm} 
\caption{(Color online) The time dependence of densities of tumor cells $x_t(t)$  calculated for $d=1, \ \Delta=3$, and several values of $p$. At the critical point ($p=0.1764$), the power-law decay $x_t\sim t^{-\alpha}$ is observed with the exponent $\alpha=0.58(1)$. For $p=0.1765$, 0.1764, and 0.1763, the simulations were made for systems with $N=10^8$ sites, and for other values of $p$, systems of size $N=10^7$ were used.
\label{time}}
\end{figure}

\begin{figure}
\includegraphics[width=9cm]{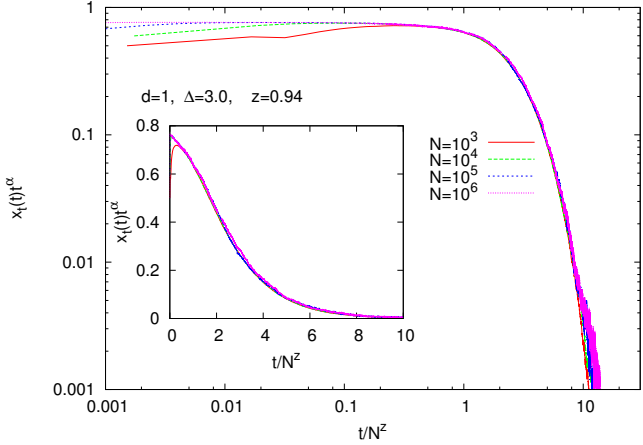}  \vspace{-0.7cm} 
\caption{(Color online) The rescaled time dependence of tumor cells density $x_t(t)$  calculated at the critical point $p=p_c=0.1764$ for several values of $N$. For $z=0.94$, the curves for different values of $N$ collapse on a single curve. The inset shows the same plot on a normal (non-logarithmic) scale. ($\alpha=0.58$) 
\label{scalingz}}
\end{figure}

\begin{figure}
\includegraphics[width=9cm]{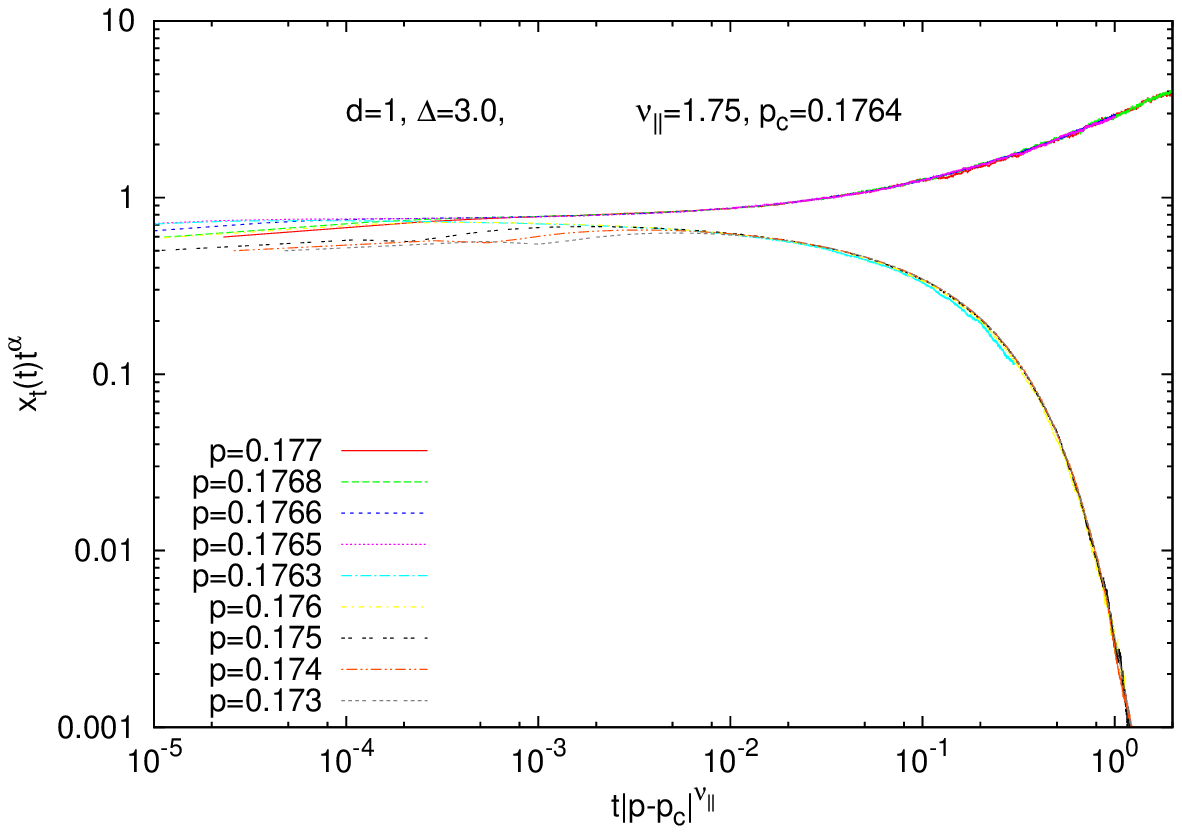}  \vspace{-0.7cm} 
\caption{(Color online) The rescaled time dependence of tumor cells density $x_t(t)$  calculated for several values of~$p$. For $\nu_{||}=1.75(5)$, the curves for different values of $p$ collapse on a single curve. ($\alpha=0.58$)
\label{scalingbeta}}
\end{figure}

The numerical results reported in the present paper  confirm the earlier findings concerning the critical behaviour of this model. In particular, more extensive simulations (by an order of magnitude both in simulation time and system size $N$) yielded a more accurate estimation of the exponent  $\alpha=0.58(1)$  (Fig.\ref{time}). 
From the finite-size scaling of $x_t(t)$ at criticality, we estimate $z=0.94(2)$ (Fig.\ref{scalingz}). With a similar procedure, the exponent $\nu_{||}$ can be  estimated from the off-critical steady-state data (Fig.\ref{scalingbeta}). The best collapse is obtained for $\nu_{||}=1.75(5)$. 


\begin{figure}
\includegraphics[width=9cm]{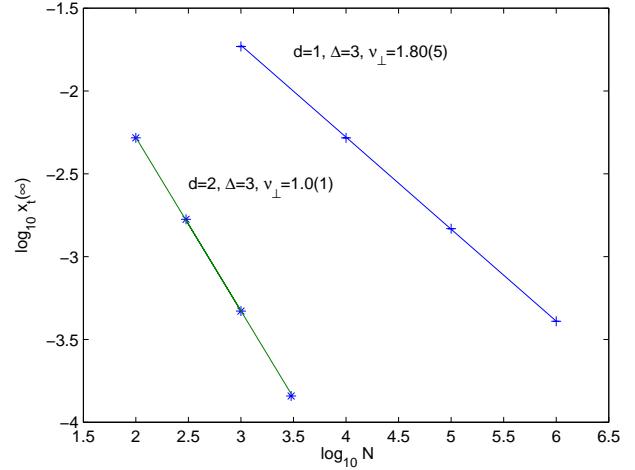}  \vspace{-0.7cm} 
\caption{(Color online) The size dependence of steady-state tumor cells density $x_t$ (averaged only over surviving runs)  calculated at the critical point of $d=1$ and $d=2$ versions of the model. From the slope of the data and relation (\ref{nu-perp}), we estimate $\beta/\nu_{\perp}$.  
\label{nuperp}}
\end{figure}

While the exponent $\nu_{||}=1.75(5)$ is quite close to its DP counterpart $\nu_{||DP}=1.73(1)$, the exponent $z=0.94(2)$ clearly differs from $z_{DP}=1.58(1)$ \cite{hinrichsen}. 
Let us also notice that from the scaling relation (\ref{alphabeta}) and our estimation of $\alpha$ and $\nu_{||}$, we obtain $\beta=1.01(5)$ and such a value agrees with the previous estimation $\beta=1.0(1)$ \cite{ferliplip} based on the steady-state values of the order parameter $x_t$ (estimations of $\beta$ based on the scaling relation (\ref{alphabeta}) usually are considered to be more accurate than those based on the steady-state data). Using the estimated value of $\beta$, from the finite-size behaviour of the tumor density at the critical point~(\ref{nu-perp}), we estimate $\nu_{\perp}=1.80(5)$ (Fig.\ref{nuperp}), which is very different from the Directed Percolation value $\nu_{\perp DP}=1.097(1)$ \cite{hinrichsen}. Within the error bars, the obtained exponents $z$, $\nu_{||}$, and $\nu_{\perp}$ satisfy the relation (\ref{znunu}).

\begin{figure}
\includegraphics[width=9cm]{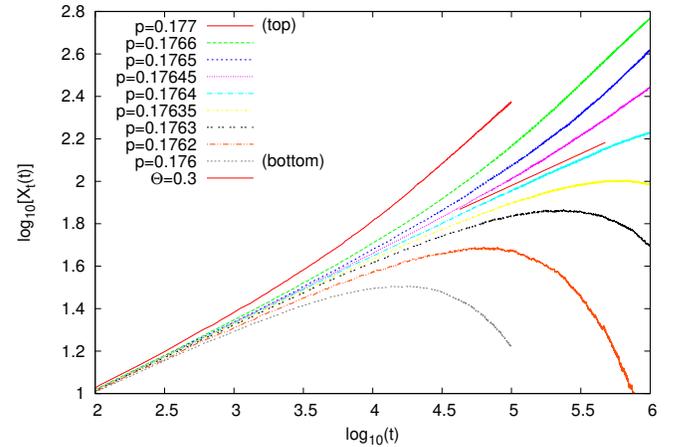}  \vspace{-0.7cm} 
\caption{(Color online) The time dependence of the number of tumor cells $X_t(t)$  calculated for $d=1$, $\Delta=3$, and several values of $p$. The initial configuration contained a single tumor cell ($X_t=1$) and all sites filled with nutrients. From numerical data for $p=0.1764$, we estimate that $\Theta=0.30(1)$, which is very close to the $d=1$ DP value $\Theta_{DP}=0.313(1)$.
\label{seed3-theta}}
\end{figure}

\begin{figure}
\includegraphics[width=9cm]{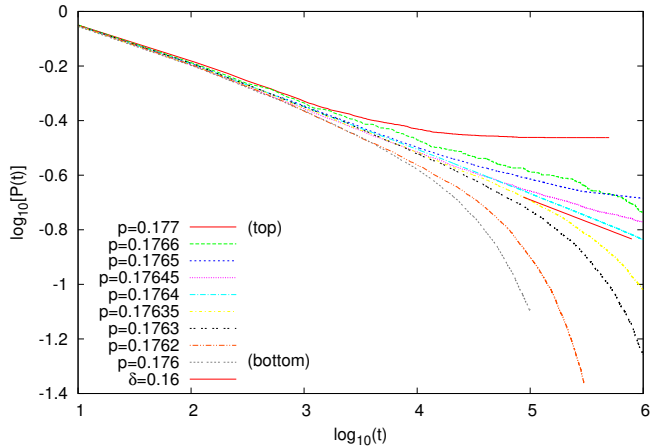}  \vspace{-0.7cm} 
\caption{(Color online) The time dependence of the tumor-survival probability $P(t)$  calculated for $d=1$, $\Delta=3$, and several values of $p$. The initial configuration contained a single tumor cell and all sites filled with nutrients. From numerical data for $p=0.1764$, we estimate that $\delta=0.160(2)$, which is very close to the $d=1$ DP value $\delta_{DP}=0.159(1)$\cite{hinrichsen}.
\label{seed3-delta}}
\end{figure}

\begin{figure}
\includegraphics[width=9cm]{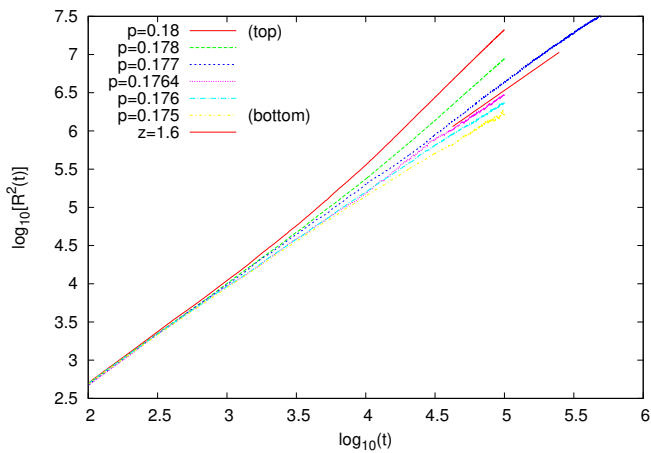}  \vspace{-0.7cm} 
\caption{(Color online) The time dependence of the mean square spreading $R^2(t)$  calculated for $d=1$, $\Delta=3$, and several values of $p$. The initial configuration contained a single tumor cell and all sites filled with nutrients. From numerical data for $p=0.1764$, we estimate that $z'=1.60(2)$, which is very close to the DP value $z_{DP}'=1.58(1)$\cite{hinrichsen}.
\label{seed3-z}}
\end{figure}

\begin{figure}
\includegraphics[width=9cm]{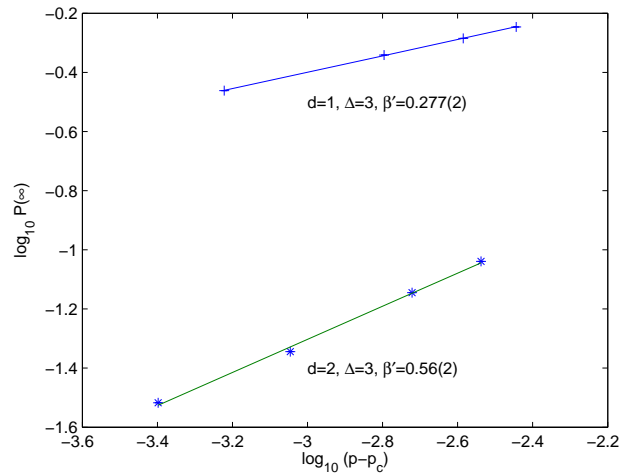}  \vspace{-0.7cm} 
\caption{(Color online) The critical decay of the ultimate survival probability $P(t\rightarrow\infty)$. Both in $d=1$ and $d=2$, the exponent $\beta'$ (estimated from the slope of the data)  is consistent with the Directed Percolation values.
\label{figBetaPrime_new}}
\end{figure}

The spreading-dynamics simulations confirm the power-law behaviour (\ref{spreading}) with $\Theta=0.30(1)$ (Fig.\ref{seed3-theta}), $\delta=0.160(2)$ (Fig.\ref{seed3-delta}) and $z'=1.60(2)$ (Fig.\ref{seed3-z}). Such values are very close to the Directed Percolation exponents $\Theta_{DP}=0.313(1)$, $\delta_{DP}=0.159(1)$ and $z'=1.58(1)$ \cite{hinrichsen}. Moreover, from the decay of $P(t\rightarrow\infty)$ (\ref{betaprime}), we estimate $\beta'=0.277(2)$ (Fig.\ref{figBetaPrime_new}), which is in a good agreement with the DP value. One can also check that $\beta'$, $\delta$, and $\nu_{||}$ satisfy the scaling relation (\ref{alphabeta}). 

The obtained results are in our opinion surprising. While some exponents ($\nu_{||}$, $\Theta$, $\delta$, $\beta'$, and $z'$) are very close to their DP counterparts, some others ($\beta$, $\alpha$, $\nu_{\perp}$, and $z$) are clearly different. 
Particularly puzzling is the fact that $z\neq z'$. While from the finite-size scaling at criticality (Fig.\ref{scalingz}) we obtain $z=0.94(2)$, the spreading dynamics (Fig.\ref{seed3-z}) leads  basically to the DP value $z'=1.60(2)$. To our knowledge, in up to now examined models with a single absorbing state those exponents are equal (although in some models, which, strictly speaking, lack an absorbing phase, they were also reported to be different \cite{nora2003}).

Using our estimations of critical exponents one can check that they satisfy the generalized hyperscaling relation (\ref{ghyper}) but only when spreading exponent $z'$ is replaced with $z=0.94(2)$ estimated from the finite-size scaling analysis.   
A further verification of scaling assumptions leading to relations (\ref{hyper})-(\ref{ghyper}) would be needed to explain such a behaviour of our model.

\subsection{d=2}
\begin{figure}
\includegraphics[width=9cm]{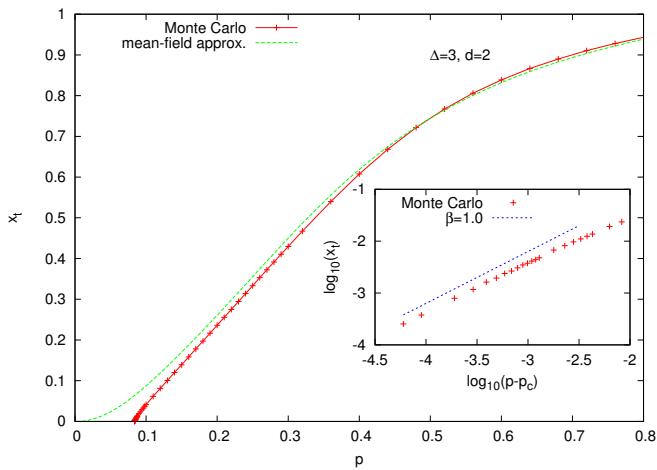}  \vspace{-0.7cm} 
\caption{(Color online) The steady-state density of tumor cells $x_t$  as a function of $p$ calculated using Monte Carlo simulations and the mean-field approximation (\ref{mfa}) for $d=2, \ \Delta=3$. The inset shows that at the critical point $p=p_c=0.08371$ the density $x_t$ decays linearly ($\beta=1)$.
\label{steadyd2}}
\end{figure}

Further numerical simulations showed that similar critical behaviour is observed in the $d=2$ version of our model. Indeed, the simulations showed that the decay of the steady-state tumor cell density $x_t$ seems to be characterized by $\beta=1.0$ (Fig.\ref{steadyd2}). Let us notice that a mean-field approximation (\ref{mfa}) remains in a reasonably good agreement with Monte Carlo data despite a low dimensionality ($d=2$) of the model.

\begin{figure}
\includegraphics[width=9cm]{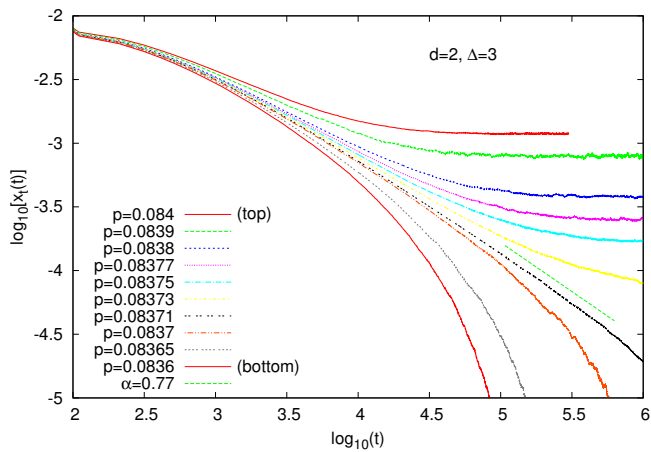}  \vspace{-0.7cm} 
\caption{(Color online) The time dependence of densities of tumor cells $x_t(t)$  calculated for $d=2, \ \Delta=3$, and several values of $p$. At the critical point ($p=0.08371$), the power-law decay $x_t\sim t^{-\alpha}$ is observed with the exponent $\alpha=0.77(1)$. 
\label{timed2}}
\end{figure}

Moreover, from the scaling relation $\beta=\nu_{||}\alpha$ and the known estimate for the $d=2$ Directed Percolation \cite{hinrichsen} $\nu_{||}=1.30(1)$, we obtain $\alpha=0.77$. The time-dependent simulations confirm that at criticality ($p=p_c=0.08371$) the decay of tumor cells follows the power law $x_t(t)\sim t^{-\alpha}$  with $\alpha\approx 0.77$ (Fig.\ref{timed2}).

\begin{figure}
\includegraphics[width=9cm]{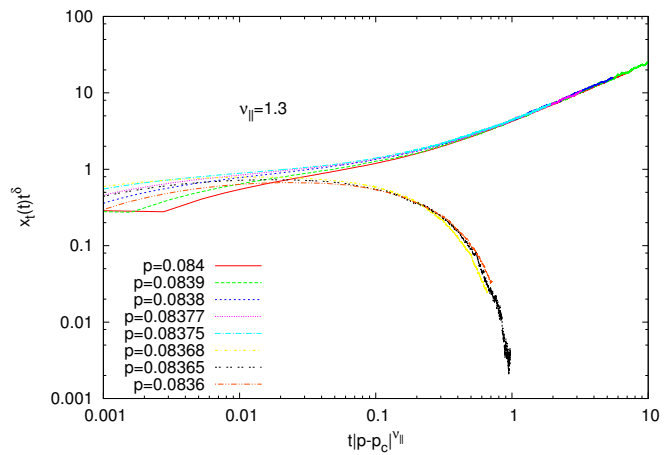}  \vspace{-0.7cm} 
\caption{(Color online) The rescaled time dependence of tumor cells density $x_t(t)$  calculated for $d=2$, $\Delta=3$, and several values of $p$. For $\nu_{||}=1.30(5)$, the curves for different values of $p$ collapse on a single curve. ($\alpha=0.77$)
\label{scalingbetad2}}
\end{figure}

From the finite-size scaling of the off-critical behaviour of $x_t(t)$, we conclude that the exponent $\nu_{||}$ is indeed very  close to the expected DP value 1.3 (Fig.\ref{scalingbetad2}). Moreover, from the finite-size behaviour of the tumor density at the critical point (\ref{nu-perp}) and taking $\beta=1.0 $, we estimate $\nu_{\perp}=1.0(1)$ (Fig.\ref{nuperp}).

\begin{figure}
\includegraphics[width=9cm]{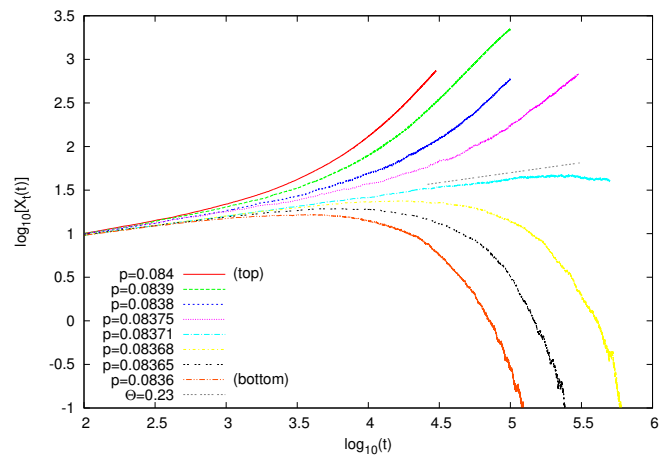}  \vspace{-0.7cm} 
\caption{(Color online) The time dependence of the number of tumor cells $X_t(t)$  calculated for $d=2$, $\Delta=3$, and several values of $p$. The initial configuration contained a single tumor cell ($X_t=1$) and all sites filled with nutrients. The simulations show that at criticality ($p=0.08371$) the exponent $\Theta$  is very close to the DP value $\Theta_{DP}=0.23(1)$.
\label{seedxtd2}}
\end{figure}

\begin{figure}
\includegraphics[width=9cm]{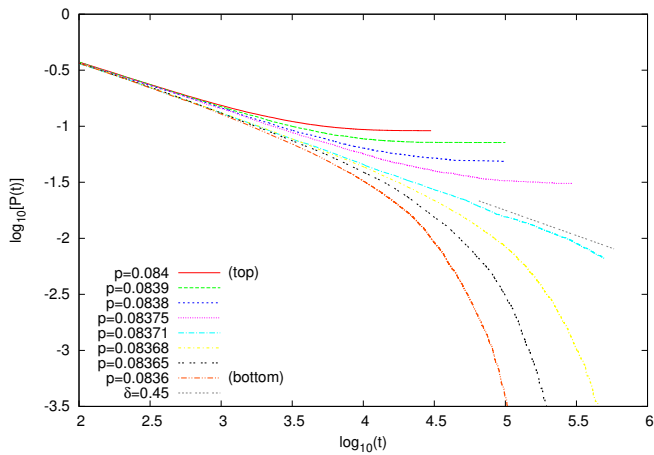}  \vspace{-0.7cm} 
\caption{(Color online) The time dependence of the tumor-survival probability $P(t)$  calculated for $d=2$, $\Delta=3$, and several values of $p$. The initial configuration contained a single tumor cell and all sites filled with nutrients. At criticality ($p=0.08371$), the exponent $\delta$  is very close to the DP value $\delta_{DP}=0.45(1)$.\cite{hinrichsen}.
\label{seedptd2}}
\end{figure}

We also used the spreading method. The obtained results, as in the $d=1$ version, are consistent with $d=2$ Directed Percolation. In particular, $\Theta$ seems to be very close to 0.23 (Fig.\ref{seedxtd2}) and $\delta\approx 0.45$ (Fig.\ref{seedptd2}).
The numerical data for the mean  square spreading are not presented here but they are also consistent with $d=2$ Directed Percolation. From the decay of $P(t\rightarrow\infty)$ (\ref{betaprime}), we estimate $\beta'=0.56(2)$ (Fig.\ref{figBetaPrime_new}), which again is in a good agreement with the DP value. All three exponents $\beta'$, $\delta$, and $\nu_{||}$ are very  close to the DP ones and they satisfy the scaling relation (\ref{alphabeta}). 

\begin{figure}
\includegraphics[width=9cm]{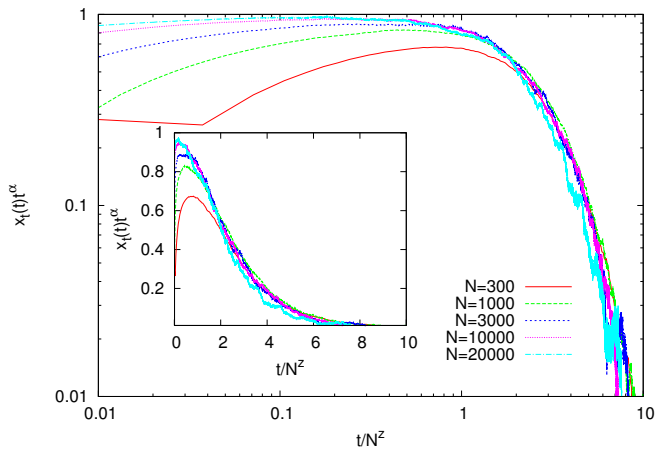}  \vspace{-0.2cm} 
\caption{(Color online) The rescaled time dependence of tumor cells density $x_t(t)$  calculated at the critical point $p=p_c=0.08371$ for $d=2$, $\Delta=3$, and several values of $N$. For $z=1.4$, the curves for different values of $N$ collapse on a single curve. The inset shows the same plot on a normal (non-logarithmic) scale. ($\alpha=0.77$) 
\label{scalingzd2}}
\end{figure}

Moreover, we estimated the exponent $z$ from the collapse of the time dependent $x_t(t)$ at criticality  and for various values of $N$ (Fig.\ref{scalingzd2}).
Let us notice that our estimate $z=1.4(1)$ (finite-size scaling), together with $\alpha=0.77(1)$, $\delta=0.45(1)$, and $\Theta=0.23(1)$, satisfies the hyperscaling relation (\ref{hyper}). Such a value of $z$ is also marginally consistent with the relation $z=\nu_{||}/\nu_{\perp}$ (\ref{znunu}).

Finally, we would like to comment on the fact that both in the $d=1$ and $d=2$ versions of our model, the exponent $\beta$ most likely takes a unit value that is characteristic to infinite-dimensional (mean-field) models with absorbing state(s). In our opinion, it might be related to a roulette-wheel selection used to choose the site for placing a nutrient. This frequently used, e.g., in genetic algorithms, selection mechanism is in a sense  nonlocal. For example, the removal of a tumor from a certain site changes the weight of this site, which affects the selection probabilites of all sites. This is because in the roulette-wheel method the selection probability of a certain site is given as a ratio of the weight of this site and the total sum of all weights (of all sites). Thus a change at a certain site changes the total sum and that  affects all the sites. 
Perhaps investigation of similar models but with different selection mechanisms would be needed to check whether the roulette-wheel selection is indeed responsible for such a critical behaviour.
But even if the critical behaviour  is changed due to the non-locality of roulette-wheel selection, some open questions still remain: (i) why only $\beta$ acquires the mean-field value; (ii) why some others exponents ($\alpha$, $z$, and $\nu_{\perp}$) are changed but not to the  mean-field values; (iii) and why some others ($\nu_{||}$, $\Theta$, $\delta$, $\beta'$, and $z'$) are intact and get the DP values.
 
\subsection{d=3}
\begin{figure}
\includegraphics[width=9cm]{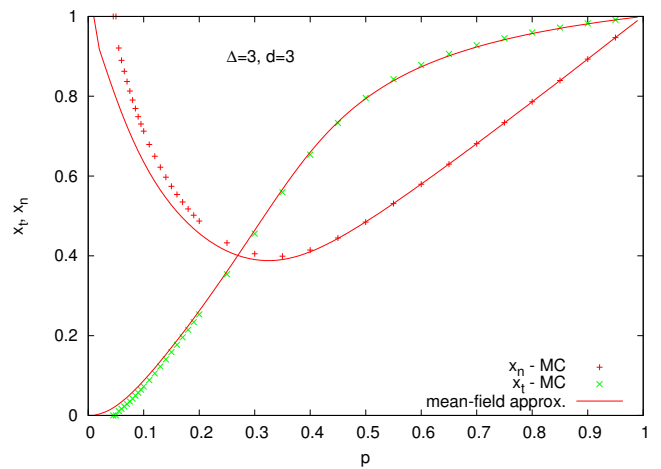}  \vspace{-0.7cm} 
\caption{(Color online) The steady-state density of tumor cells $x_t$  and of nutrients $x_n$ as a function of $p$ calculated using Monte Carlo simulations and the mean-field approximation (\ref{mfa}) for $d=3, \ \Delta=3$ and $N=100$.
\label{steadyd3}}
\end{figure}

As a last of our results, we present the steady state behaviour of our model in the $d=3$ version. Qualitative dependence of $x_n$ and $x_t$ on $p$ (Fig.\ref{steadyd3}) is similar to lower-dimensional versions. As expected, one can notice a much better agreement with mean-field approximation (\ref{mfa}). However, one can also observe that region of existence of the absorbing phase ($x_n=1,\ x_t=0$) shrinks. It means that even for low nutrient update rate $p$ tumor cells survive in the system. Thus starving tumor to death is less effective in $d=3$ systems but unfortunately this is the dimension that is most relevant from biological and clinical point of view. 

\section{Conclusions}
We presented a more detailed analysis of the critical behaviour of a recently introduced tumor growth model. Although the model has a single absorbing state, its critical behaviour differs from the expected Directed Percolation universality class in an intriguing way  In particular, only some of its critical exponents, namely $\beta$, $\alpha$, $\nu_{\perp}$, and $z$, differ from their DP counterparts. Other exponents, namely $\nu_{||}$, $\beta'$, and spreading exponents $\Theta$, $\delta$, and $z'$, most likely take the  DP values. Moreover, the generalized hyperscaling relation (\ref{ghyper}) seems to be satisified in our model, provided, however, that the spreading exponent $z'$ is replaced with the dynamical exponent $z$.

What is also interesting is the fact that in our model the exponent $\beta$ seems to take the mean-field value $\beta=1$. In a speculative way, we attribute this fact to, in a sense, non-locality of the roulette-wheel selection, which is used in our model to choose a site to supply a nutrient. Even if it is so, it still remains unclear why such a non-locality affects only   some of the exponents.

Let us also mention that the fact that $\beta=1$ exactly (if proven) might be used to precisely locate the critical point from the decay of the order parameter. With a precisely located critical point, one might expect that also other critical exponents, including DP ones, will be determined accurately. 

We hope that clarifying the behaviour of the present model might contribute to a better understanding of Directed Percolation universality class, which is one of the most basic among nonequilibirum critical phenomena.

Various models of  tumor growth processes were used and shown to provide a valuable support to biological and clinical studies \cite{byrne}.  Often these models are very complex and  their understanding  is limited.   Our approach uses  a simplified model where more detailed understanding seems to be feasible. It would be interesting to examine more realistic extensions of our model that takes into account for example remodelling of vascular network that accompanies a tumor growth process\cite{rieger}.



\begin{thebibliography}{}
\bibitem{comment1} In our previous paper \cite{ferliplip}, this exponent was denoted by $\delta$, which is, however, often used to denote the exponent describing the decay of the surivival probability $P(t)\sim t^{-\delta}$. Although typically $\delta=\alpha$, for our model these exponents are very different.
\end{thebibliography}
\end {document}